\documentstyle[prl,aps,preprint]{revtex}

\begin{document}
\draft
\preprint{}
\title{\bf An Exactly Solvable Model of Fermions with Disorder.}
\author{A. M. Tsvelik}
\address{Department of Physics, University of Oxford, 1 Keble Road,\\
Oxford,OX1 3NP, UK}
\date{\today}
\maketitle

\begin{abstract}

 Non-perturbative results are obtained for
multi-point correlation functions of
the model of (2 + 1)-dimensional relativistic fermions
in   a random non-Abelian gauge
potential. The results indicate
that the replica symmetry  for this model is unbroken. We calculate
the
diffuson
propagator and show that
DC-conductivity for this model  is
finite.

\end{abstract}

\narrowtext

 In this letter we continue to study
the model of (2 + 1)-dimensional  relativistic
fermions interacting with random non-Abelian gauge potential. This
model was  introduced in Ref. 1  in relation with effects of
disorder in two dimensional d-wave
superconductors. It was shown that
after averaging over the static disorder  one
gets the following Euclidean action:
\begin{eqnarray}
S = S_0 + \int d^2x[ i\epsilon\bar\psi_{\alpha,p}\psi_{\alpha,p} +
\frac{i\omega + 0}{2}\bar\psi_{\alpha,p}\Lambda^3_{pq}\psi_{\alpha,q}]\\
S_0 = \int
d^2x[\bar\psi_{\alpha,p}\gamma_{\mu}\partial_{\mu}\psi_{\alpha,p} +
cJ^a_{\mu}J^a_{\mu}]
\end{eqnarray}
Spinor fields $\psi_{\alpha,p}$ with $p = 1, ... R$
and $p = R + 1, ... 2R$ describe Fourier components of the original
fermions with frequencies $\epsilon + \omega/2$ and  $\epsilon -
\omega/2$ respectively, where $R \rightarrow 0$ is the number of
replicas. Greek indices run from 1 to $N$.
The flavour currents are given by
\[
J^a_{\mu} =
\sum_{p =
1}^{2R}\bar\psi_{\alpha,p}\gamma_{\mu}\tau^a_{\alpha\beta}\psi_
{\beta,p}
\]
with $\tau^a$'s being generators of the SU(N) group. In the context
of
disordered superconductor $N$ denotes
the number of nodes of the order parameter on the Fermi surface. The coupling
constant $c$ is related to the random potential. The matrix
$\Lambda^3$ is the diagonal matrix with $R$ elements $+1$ and $R$
elements $-1$ on the
main diagonal: $\Lambda^3 = diag(1,...1;- 1, ... -1)$. In what follows
we shall also use the matrix $\Lambda^1$
whose  only nonzero elements are situated on the
main antidiagonal: $\Lambda^1 = antidiag(1, 1, ...1)$.

 The Abelian version of this model has been studied by Ludwig et.
al$^2$.
The density of states (DOS) is expressed as
\begin{equation}
\rho(\epsilon) = \frac{1}{\pi}\Re e\lim_{R\rightarrow
0}R^{-1}Tr<\bar\psi(x)\psi(x)>|_{\omega = 0}
\end{equation}
The main difference between the present  problem
and the conventional localization is that in the present case DOS
 is strongly affected by the disorder.
Being linear in energy in the absence of
disorder ($\rho(\epsilon) \sim |\epsilon|$), DOS changes its behavior
at low energies  $\rho(\epsilon) \sim |\epsilon|^{\nu}$
where $\nu = 1/(2N^2 - 1)$$^1$. It will be shown later that with
DOS vanishing at zero energy the low energy states remain delocalized.
Another difference is that the conductivity of
(2 + 1)-dimensional massless Dirac
fermions with an arbitrary  dumping is constant$^{2,3}$:
\begin{equation}
\sigma_{xx} = Ne^2/2\pi^2\hbar \label{eq:sig}
\end{equation}
The incensitivity of the conductivity to dumping makes one suspect
that it may remain finite in the presence of
disorder. We shall prove that this is indeed
the case.

 The model (1,2), as it stands,
contains fast and slow degrees of freedom. The
latter are $SU(N)$-singlets described by the well known
$Q$-matrices ($2R \times
2R$-matrices). The symmetry of the model dictates that $Q$-matrices
belong to the $U(2R) = U(1)\times SU(2R)$ group.
The integration over the fast degrees of freedom performed
in Ref. 1 gives the effective action for the matrices
$Q = g\exp[i\gamma\Phi]$ ($\gamma = \sqrt{2\pi/NR}$)
in the form of the Wess-Zumino
model on the group $U(1)\times SU(2R)$:
\begin{eqnarray}
S = S_0 + M\epsilon Tr(Q + Q^+) + iM\frac{\omega}{2}Tr[\Lambda(Q +
Q^+)] \nonumber\\
S_0 = \frac{1}{2}\int d^2x (\partial_{\mu}\Phi)^2 + NW[SU(2R); g]\nonumber\\
W[SU(2R); g] = \frac{1}{16\pi}\int d^2x
Tr(\partial_{\mu}g^+\partial_{\mu}g) + \frac{1}{24\pi}\int d^3x
\epsilon_{abc}
Tr(g^+\partial_{a}g g^+\partial_b g g^+\partial_c g) \label{eq:wzw}
\end{eqnarray}
  The quantity $M$ is the energy
scale introduced by the disorder $M \sim \exp[ - 2\pi/Nc]$,
which marks the crossover from the bare DOS
$\rho(\epsilon) \sim |\epsilon|$ to  the renormalized
DOS $\rho(\epsilon) \sim |\epsilon|^{\nu}$. In this
Wess-Zumino  model $M$ serves as  the ultraviolet
cutoff. It is worth remarking  that the
Wess-Zumino  action {\it is not equivalent}
to the action with the
topological term
which describes  the conventional localization problem
in a strong magnetic field.

 The Wess-Zumino model is well studied at  finite $R$, where it
is  a critical theory with power law decay of correlation
functions.
Our primary objective is  to show that this model remains well defined
in
the replica limit, that is its correlation functions have finite
values at $R \rightarrow 0$.  The closest
implication of this fact is that the replica symmetry remains
unbroken.

 In
determining the replica limit we follow the
general principle: calculating
any N-point correlation function $F_{2R}(1, 2, ...N)$ one treats
$R$
as an arbitrary $integer$ number on all intermediate steps
of calculations until the finite expression is
obtained. The replica limit is then defined as follows:
\begin{equation}
F(1,2, ... N) = \lim_{R \rightarrow 0}\frac{1}{2R}F_{2R}(1, 2, ...N)
\label{eq:def}
\end{equation}

 Let us study the correlation functions of the $Q$-fields. The problem
of indices is simplified
by the fact that $Q_{pq}$ matrices are slow parts of the
operators
\begin{equation}
Q_{pq} \sim \sum_{\alpha = 1}^N \psi^+_{R,\alpha,p}\psi_{L,\alpha,q}
\end{equation}
This fact provides us with a simple recipe: any
$N-$point correlation function has the same index structure as  the
$N$-point function of the
$\sum_{\alpha}\psi^+_{R,\alpha,p}\psi_{L,\alpha,q}$-fields
in  the theory of
massless free fermions. The simpliest example
is the 2-point function:
\begin{eqnarray}
<Q_{p_1r_1}(z,\bar z)Q^+_{r_2p_2}(0,0)> =
\delta_{r_1r_2}\delta_{p_1p_2}\frac{1}{(M|z|)^{4\Delta_1}} \label{eq:corr}
\end{eqnarray}
where
$\Delta_1$ is the conformal dimension of the composite
operator $Q$ given by the sum of the dimensions of the bosonic
exponent $\exp[i\sqrt{2\pi/NR}\Phi]$ and the operator
field  $g_{pr}$ from the
fundamental representation of the SU(2R) group. Using the results of
Ref. 4 we get:
\begin{equation}
\Delta_1 = \lim_{R \rightarrow 0}[\frac{1}{4RN} + (2R - 1/2R)/(N +
2R)] = \frac{1}{2N^2}  \label{eq:dimen}
\end{equation}
In the replica limit we get from (~\ref{eq:corr})
\begin{equation}
D(z, \bar z) \equiv \lim_{R\rightarrow 0}R^{-1}<Tr[\Lambda^1 Q(z, \bar z)]
Tr[\Lambda^1 Q^+(0,0)]> = (M|z|)^{- 2/N^2}
\end{equation}
We identify this correlation function
with the propagator of diffuson in the zero
frequency limit. Its Fourier transformation gives the propagator  in
momentum space:
\begin{equation}
D(\omega = 0, q) \sim \frac{D_0(q/M)^{2/N^2}}{q^2}
\end{equation}
where $D_0$ is the bare diffusion constant.

 We can identify other operators and their conformal
dimensions. Thus, the operator expansion of two $Q$ matrices contain
the symmetric and antisymmetric fields $O_S$ and $O_A$:
\begin{eqnarray}
Q(1)Q(2) =
\exp[i\sqrt{2\pi/RN}\Phi(1)]g(1)\exp[i\sqrt{2\pi/RN}\Phi(2)]g(2) \sim
\nonumber\\
\exp[i\sqrt{8\pi/RN}\Phi(1)]O_S(1) +
\exp[i\sqrt{8\pi/RN}\Phi(1)]O_A(1) + ...
\end{eqnarray}
where dots stand for nonsingular terms. Taking
the corresponding conformal dimensions from  \cite{Knizh84} we get the
conformal dimensions of the composite fields $\tilde O_{S,A} =
\exp[i\sqrt{8\pi/RN}\Phi]O_{S,A}$ in the replica limit:
\begin{eqnarray}
\Delta_A = \lim_{R\rightarrow 0}[\frac{1}{RN} + \frac{(R - 1)(2R +
1)}{R(N + 2R)}] = \frac{2 - N}{N^2}\nonumber\\
\Delta_S = \lim_{R\rightarrow 0}[\frac{1}{RN} + \frac{(R + 1)(2R -
1)}{R(N + 2R)}] = \frac{2 + N}{N^2} \label{eq:dim}
\end{eqnarray}
We see that that the scaling dimension of the antisymmetric field
is negative for $N > 2$, which indicates  that the replica limit of the
model (1,2) is a nonunitary theory. The next field is the adjoint
field $O_{ad}$ appearing in the operator expansion of $QQ^+$:
\begin{equation}
Q(1)Q^+(2) \sim g(1)g^+(2) \sim Tr[\tau^ag^+\tau^bg] \equiv
O_{ad}^{ab}
\end{equation}
According to Ref. 6  the corresponding conformal dimension is
$\Delta_{ad} = 2R/(2R + N)$. It  vanishes in the replica limit and,
as we shall
see later, this field just does not exist in the replica limit.

 The central charge of our theory is the sum of central charges
of the free bosonic field ($C = 1$) and  the
Wess-Zumino-Witten model on the $SU(2R)$ group:
\begin{equation}
C = 1 + \frac{N(4R^2 - 1)}{N + 2R} = \frac{2R}{N} + O(R^2)
\end{equation}
Thus the resulting
central charge vanishes; however,
according to the definition of the replica limit
(~\ref{eq:def})
the  physical correlation function
of the stress-energy tensors does not vanish:
\begin{equation}
<T(z)T(0)> = \lim_{R\rightarrow 0}\frac{C_R}{2R}\frac{1}{2z^4} =
\frac{1}{2Nz^4}
\end{equation}

 Let us now study the four point correlation function of the $Q$
fields and show that it has the correct replica limit. For this
purpose we use the result obtained in Ref. 4:
\begin{eqnarray}
<Q_{p_1r_1}(z_1,\bar z_1)Q^+_{r_2p_2}(z_2,\bar z_2)Q_{p_3r_3}
(z_3,\bar z_3)Q^+_{r_4p_4}(z_4, \bar z_4)> = \nonumber\\
\left[\frac{|z_{14}z_{23}|}{|z_{12}z_{14}z_{13}z_{24}|}\right]^{2/N^2}(F
+ \tilde F),\\
\tilde F =
[\delta_{p_1p_2}\delta_{p_3p_4}\delta_{q_1q_2}\delta_{q_3q_4}G_{11}(x,\bar
x) + \delta_{p_1p_4}\delta_{p_3p_2}\delta_{q_1q_4}\delta_{q_3q_2}G_{22}(x,\bar
x)],\\
F = [\delta_{p_1p_2}\delta_{p_3p_4}\delta_{q_1q_3}\delta_{q_2q_4}
G_{12}(x,\bar
x) + \delta_{p_1p_4}\delta_{p_3p_2}\delta_{q_1q_2}\delta_{q_3q_4}G_{21}(x,\bar
x)]
\end{eqnarray}
where
\[
x = \frac{z_{12}z_{34}}{z_{14}z_{32}}, \bar x = \frac{\bar z_{12}\bar
z_{34}}{\bar z_{14}\bar z_{32}}
\]
and the functions $G_{AB}, A,B = 1,2$ are known functions. Now note
that in our theory we shall deal only with correlation functions of
$TrQ$, $Tr\Lambda^3 Q$ and $Tr\Lambda^1Q$.
Since all correlation functions must be
proportional to $R$, only the term with $F$ (that is the term with
all indices equal) survives
in the replica limit,
the term with $\tilde F$ being proportional $R^2$.
Therefore we have
%\begin{eqnarray}
%D(1,2,3,4) = <TrQ(1)TrQ^+(2)TrQ(3)TrQ^+(4)> =
%<Tr\Lambda Q(1)Tr\Lambda Q^+(2)
%Tr\Lambda Q(3)Tr\Lambda Q^+(4)> =  \nonumber\\
%<Tr\Lambda Q(1)Tr\Lambda Q^+(2)
%TrQ(3)TrQ^+(4)> =  <TrQ(1)TrQ^+(2)
%Tr\Lambda Q(3)Tr\Lambda Q^+(4)> =  \nonumber\\
%<Tr\Lambda Q(1)TrQ^+(2)
%TrQ(3)Tr\Lambda Q^+(4)> =  <Tr\Lambda Q(1)TrQ^+(2)
%Tr\Lambda Q(3)TrQ^+(4)>
%\end{eqnarray}
%where
\begin{eqnarray}
R^{-1}<Tr\Lambda^1 Q(1)TrQ^+(2)TrQ(3)Tr\Lambda^1 Q^+(4)> =
\left[\frac{|z_{14}z_{23}|}{|z_{12}z_{14}z_{13}z_{24}|}\right]^{1/N^2}\times
\nonumber\\
\left[\bar x F(1/N,
- 1/N, 1, x)F(1 + 1/N, 1
- 1/N, 2, \bar x) + (x \rightarrow \bar x)\right]
\end{eqnarray}

 In order to get the operator algebra of the model we consider
various limits of this formula. In the limit $z_{12} = z_{34} = \delta
\rightarrow 0$ we get
\begin{equation}
D(1,1 + 0; 3,3 + 0) = (\delta)^{-4/N^2} + \frac{1}{z_{13}^2} +
\frac{1}{\bar z_{13}^2}
\end{equation}
{}From this we derive the following operator expansion:
\begin{equation}
Q(1)Q^+(2) \sim |z_{12}|^{-2/N^2}(I + z_{12}J(2) + \bar z_{12}\bar
J(2) + ...)
\end{equation}
where $J, \bar J$ are the left and right current operators. Notice
that the primary operator in the adjoint representation present
for $R \neq 0$ does not appear in this expansion.

In the limit $z_{13} = z_{24} = \delta \rightarrow 0$ we have
\begin{equation}
D(1, 2;1 + 0, 2 + 0) = |z|^{4(N - 2)/N^2}C_1 + |z|^{- 4(N + 2)/N^2}C_2
\end{equation}
($C_1, C_2$ are constants)
which shows the presence of the symmetric and antisymmetric fields
with conformal dimensions (~\ref{eq:dim}) in the operator
expansion of $QQ$ and $Q^+Q^+$. These operators describe mesoscopic
fluctuations of the local density of states.

 Now let us discuss a crossover to small momenta. As follows
from Eq.(~\label{eq:dimen}) the operators $Tr(1 \pm \Lambda^3)Q$
are relevant perturbations with scaling dimensions $2 - 2\Delta_1 = 2
- 1/N^2$. Therefore at finite $\omega$
one can conjecture the following form for the diffuson propagator:
\begin{equation}
D_{\epsilon = 0}(\omega,q) = \frac{1}{q^{2 - 2/N^2}}
f\left(\frac{\omega^2}{q^{4 - 2/N^2}}\right) \label{eq:dif}
\end{equation}
%The fact that $\epsilon$ and $\omega$ appear only in the combination
%$s^2 = \epsilon^2 - \omega^2/4$ follows from the structure of the
%perturbation series.

 Eq.(~\ref{eq:dif}) shows that  the spectrum of the diffusive modes $i\omega =
D(\omega)q^2 \sim q^{2 - 1/N^2}$ and the diffusion coefficient scales
as $D(\omega) \sim \omega^{- 1/(2N^2 - 1)} \sim \rho^{-1}(\omega)$.
That is the quantity $\rho(\omega)D(\omega)$ stays constant under
renormalization (the presence of logarithms or finite corrections
is not excluded). Since from the other
side the conductivity is proportional to
$\sigma_{xx} \sim \rho(0)D(0)$, the absence of renormalization of
$\rho(\omega)D(\omega)$ at finite frequences makes it likely that the
conductivity remains finite in the presence of weak disorder.

 We can go futher than this qualitative analysis.
Namely, using the peculiar properties of the action (~\ref{eq:wzw}) we can
calculate the propagator of diffuson
in the limit $q << |\omega|^{1/1 - \Delta_1}$. Let us
introduce  the following change into
the action (~\ref{eq:wzw}):
\begin{equation}
Tr(1 \pm \Lambda^3)(Q + Q^+) \rightarrow Tr(1 \pm \Lambda^3)(h^+Q +
Q^+h)
\end{equation}
where $h(x)$ is an external source - a matrix from the $U(2R)$ group.
It turns out that in
the limit $q << |\omega|^{1/1 - \Delta_1}$
we can integrate over $Q$-variables and get the effective action
for $h(x)$ in the form of expansion in $|\omega|^{ - 1/1 - \Delta_1}$.
Indeed, let us make a shift:
$\tilde Q = h^+Q, Q = h\tilde Q$. Using the
well known property of the Wess-Zumino action, we write
\begin{eqnarray}
S = S_0(h\tilde Q) + M\epsilon Tr(\tilde Q + \tilde Q^+) +
iM\frac{\omega}{2}Tr[\Lambda^3(\tilde Q +
\tilde Q^+)] = \nonumber\\
S(\tilde Q; \epsilon, \omega) + S_0(h) + \frac{N}{8\pi}\int d^2x
Tr[h^+\partial_{\mu}h \tilde Q\partial_{\mu}\tilde Q^+] \label{eq:rot}
\end{eqnarray}
Since fluctuations of the $\tilde Q$-field are governed by the
Wess-Zumino  action perturbed by relevant operators, they
have a finite correlation length
$\xi \sim (|\omega|)^{- 1/(1 - \Delta_1)}$. Therefore
the last term in the action (~\ref{eq:rot}) gives corrections
to $S_0(h)$ which contain powers of $\xi\nabla h$.
At small momenta such perturbations can be neglected. Thus
at small wave vectors
$S_0(h)$ effectively becomes a  generating functional of correlation functions
of
$Q$. In particular, we have
\begin{equation}
K(\epsilon = 0, \omega; q) \equiv
\frac{M^2}{2R}<Tr(1 + \Lambda^3)\Lambda^1 Q Tr(1 - \Lambda^3)\Lambda^1
Q^+> = - \frac{N}{2\pi}\frac{q^2}{\omega^2}(1 +
O(q^2/|s|^{1 - \Delta_1})) \label{eq:ans}
\end{equation}
This means that the crossover function $f(x)$ in Eq.(~\ref{eq:dif}) behaves
as
\[
f(x) = \frac{N}{2\pi}\frac{1}{x}
\]
at large x.  This result allows us to calculate the DC-conductivity
$\sigma_{xx}$. Using the definition given in  Ref. 2 we get from
(~\ref{eq:ans})
\begin{equation}
\sigma_{xx} = - \frac{e^2}{\pi\hbar}\lim_{\omega\rightarrow
0}\omega^2\left(\frac{\partial}{\partial q^2}\right)_{|q = 0}
K(0, \omega;q) = \frac{Ne^2}{2\pi^2\hbar}
\end{equation}
Comparing this result with Eq.(4) we see  that the DC-conductivity remains
unrenormalized and therefore the disorder manifest itself only at
large momenta. In the limit $q \rightarrow 0$ there are no singular
contributions. This is not typical for conventional
disordered systems where disorder changes a statistics of
low-lying  energy levels. In the same time one probably should not be
surprized by our results. After all they show that the model with
disorder remains integrable and we know that integrability and
fractality of the spectrum are inconsistent.

\acknowledgments
The author  acknowledges valuable discussions with B. Altshuler, J.
Chalker, A. Nersesyan, I. Lerner, P. Lee and S. Sondhi. I am particularly
grateful to A. Cappelli for pointing out a mistake in the original
version of the paper.

\end{document}